\documentclass[a4paper]{article}

\usepackage{url} 
\usepackage[finnish, english]{babel}
\usepackage{amsmath}
\usepackage{amssymb}
\usepackage[dvipdf]{graphicx}
\usepackage{color}
\usepackage{url}
\usepackage{epstopdf}

\usepackage{ctable}
\usepackage{epsfig}
\usepackage{subfigure}
\usepackage{float}
\usepackage[small]{caption}
\usepackage{subfig}
\usepackage{multirow}

\graphicspath{{pics_p/}}

\def\reals{\mathbb{R}}

\def\logit{\mathop{\rm logit}\nolimits}

\def\bin{\mathop{\rm Bin}}
\def\R{\mathbb{R}}
\def\E{E(x\!+\!1,t\!+\!1)}

\def\minimize{\mathop{\rm minimize}\limits}

\title{Liability-driven investment in longevity risk management}

\author{Helena Aro\footnote{Department of Mathematics and Systems Analysis, Faculty of Information and Natural Sciences, Aalto University, helena.aro@aalto.fi}
\and  Teemu Pennanen\footnote{Department of Mathematics, King's College London}
}

\date{\today}

\begin{document}

\maketitle

\begin{abstract}
This paper studies optimal investment from the point of view of an investor with longevity-linked liabilities. 
The relevant optimization problems rarely are analytically tractable, but we are able to show numerically that liability driven investment can significantly outperform common strategies that do not take the liabilities into account. 
In problems without liabilities the advantage disappears, which suggests that the superiority of the proposed strategies is indeed based on connections between liabilities and asset returns.


\end{abstract}


 \vspace{4mm}

\section{Introduction}

Longevity risk, the uncertainty in future mortality developments, affects pension providers, life insurers, and governments. The population structure of developed countries is increasingly leaning towards the old, and the effects of medical advances and lifestyle choices on mortality remain unpredictable, which creates an increasingly acute need for life insurance and pension plans to hedge themselves against longevity risk.

Various longevity-linked instruments have been proposed for the management of longevity risk; see e.g.\ \cite{BlakeBurrows, BCDMortalitySecurities, BiffisBlake, DowdSurvivorSwaps, LinCoxMBonds}. It has been shown how such instruments, once in existence, can be used to hedge mortality risk exposures in pensions or life insurance liabilities \cite{BlakeBurrows, BCDM2006, Blake2012, Cairns2011a, LiHardy2011, DMM2008}. Indeed, demand for longevity-linked instruments appears to exist, and some longevity transactions have already taken place. However, a major challenge facing the development of longevity markets is the hedging of the risk that stems from issuing longevity-linked securities. The supply for mortality-linked instruments might increase if their cash-flows could be (partially) hedged by appropriately trading in assets for which liquid markets already exist. Such a development has been seen e.g.\ in options markets, which flourished after the publication of the seminal Black--Scholes--Merton model. 
Even though the cash flows of mortality-linked instruments cannot be perfectly replicated, it may be possible to diminish the residual risk by an appropriate choice of an investment strategy. 


This paper addresses the above issues by studying optimal investment from the point of view of an insurer with longevity-linked liabilities. As such problems are rarely analytically tractable, we employ a numerical procedure that adjusts the investment strategy according to the statistical properties of assets and liability as well as a given risk measure. The approach can be applied in pricing and hedging of longevity-linked instruments, as well as in asset-liability management of pension plans and life insurers. Rather than aiming at general investment principles, we illustrate the technique in the specific example of hedging a survivor bond whose payments are tied to a given cohort over a fixed time interval.



The most straightforward approach to hedging of longevity-linked instruments is {\em natural hedging}, where an insurer hedges longevity risk by taking positions with opposite exposures to longevity developments \cite{cox2007natural, NaturalWang2010}. Such an approach is obviously limited by the demand on the relevant insurance products. Another popular approach builds on {\em risk neutral valuation} which is based on the no-arbitrage principle from financial economics; see e.g.\ \cite{BCDM2006,mdfrm,ballotta2006fair, PricingDeath} and Section~10 of  \cite{wilkie2003reserving}. In analogy with the Black--Scholes--Merton theory, it has been suggested that longevity-linked instruments could then be hedged using {\em delta hedging} by determining price sensitivities with respect to traded securities. This approach is, however, invalidated by the fact that the payouts of longevity-linked instruments cannot be replicated by liquidly traded assets as assumed by the risk neutral valuation theory; see the discussion in \cite{wilkie2003reserving,barrieu2012understanding, Bauer2010}.

This paper employs a computational technique that constructs diversified strategies from a family of simpler {\em basis strategies}. We find that the risk associated with the diversified strategy diminishes significantly when one includes basis strategies suggested by the statistical connections between mortality and financial markets observed in \cite{morfin}. To assess to which extent the reduction of risk is due to the asset-liability connections, we performed the same computations also without liabilities. In this case, the inclusion of the liability-driven investment strategies had negligible effect on risk, which suggests that the reduction of risk in the asset-liability management problem was indeed due to the connections between assets and liabilies.

The rest of this paper is organized as follows. Section 2 formulates the asset-liability management problem of a longevity-linked cash flow. Section 3 introduces investment strategies that serve as basis strategies for the computational procedure described in Section 4. Section 5 presents results from a simulation study, and Section 6 concludes.

\section{The asset-liability management problem}

Consider an insurer with given initial capital $w_0$ and longevity-linked liabilities with claims $c_t$ over time $t=1,2,\ldots,T$. After paying out $c_t$ at time $t$, the insurer invests the remaining wealth in financial markets. We look for investment strategies whose proceeds fit the liabilities as well as possible in the sense of a given risk measure $\rho$ on the residual wealth at time $T$.

We assume that a finite set $J$ of liquid assets (bonds, equities, \ldots) can be traded at $t=0,\ldots,T$. The total return on asset $j$ over period $[t-1,t]$ will be denoted $R_{t,j}$. The amount of cash invested in asset $j$ over period $(t,t+1]$ will be denoted by $h_{t,j}$. The asset-liability management problem of the insurer can then be written as
\begin{equation}\tag{ALM}
\begin{aligned}
& {\text{minimize}}\quad & \rho(\sum_{j \in J}&h_{T,j})\quad \text{over}\quad h \in \mathcal{N} \\
& \text{subject to}\quad & \sum_{j \in J}h_{0,j} &\le w_0 \\
&& \sum_{j \in J}h_{t,j} &\le \sum_{j \in J}R_{t,j}h_{t-1,j} - c_t\quad t=1,\ldots,T \\
&& h_t &\in D_t,\quad t=1,\ldots,T \\
\end{aligned}
\label{alm}
\end{equation}
The liabilities $(c_t)_{t=0}^T$ and the investment returns $(R_{t,j})_{t=0}^T$ will be modeled as stochastic processes on a filtered probability space $(\Omega,\mathcal{F},(\mathcal{F})_{t=1}^T,P)$. The set $\mathcal{N}$ denotes the $\reals^J$-valued $(\mathcal{F}_t)_{t=1}^T$-adapted investment strategies $(h_t)_{t=0}^T$. Being adapted means that the portfolio $h_t$ chosen at time $t$ may only depend on information observed by time $t$. The last constraint describes portfolio constraints. The set $D_t$ of feasible strategies is allowed to be random but known at time $t$.\footnote{More precisely, $D_t$ is assumed $\mathcal{F}_t$-measurable, i.e.\ $\{\omega\in\Omega\,|\, D_t(\omega)\cap U \ne\emptyset\}\in\mathcal{F}_t$ for every open $U$. } Short-selling constraints, for instance, correspond to the deterministic constraint $D_t(\omega)=\R^J_+$. Describing limitations on investment strategies in case of negative wealth require a random $D_t$.

The risk measure $\rho$ is a convex function on the space of real-valued random variables. It describes the insurer's preferences over random terminal wealth distributions. We refer the reader to \cite[Chapter~4]{FS} for a general treatment of risk measures. In addition to the terminal wealth, one may also wish to take into consideration the overall trajectory of wealth either in the objective or the constraints. For simplicity, we will concentrate on the above formulation, which is more in line with established models of mathematical finance; see e.g. \cite[Chapter~8]{FS} or \cite[Chapter~3]{ds6}.

{\em Liability-driven investment} refers to the general principle that optimal investment strategies depend on the liabilities. The same idea is behind the famous Black--Scholes--Merton option pricing model where the price of an option is defined as the least amount of initial capital that allows for the implementation of an investment strategy whose proceeds match the option payout exactly. In the case of longevity-linked liabilities, exact replication is not possible so one has to evoke the risk preferences as is done in problem \eqref{alm} in terms of the risk measure $\rho$.

Problems of the form \eqref{alm} arise naturally in reserving for existing insurance liabilities as well as in underwriting new insurance contracts. Optimal risk adjusted reserves are obtained as the least initial wealth $w_0$ that allow for acceptable levels of risk in \eqref{alm}. In underwriting, one looks for a premium that would allow the insurer to take on the additional liabilities without worsening the optimal value of \eqref{alm}; see \cite{pen13} for a general study of risk management-based valuation of uncertain cash-flows.



\section{Investment strategies}
\label{stra_section}

In this section we present investment strategies that are used in subsequent numerical illustrations. 
We recall some well-known trading strategies recommended for long-term investment, and also introduce new strategies that try to employ the connections between the longevity-linked liabilities and asset returns.

We will describe the proportions of wealth invested in each asset $j\in J$ at time $t$ by vector $\pi_t=[\pi^1_t, \pi^2_t,\ldots,\pi^J_t ]$ whose components sum up to one. The amount of wealth invested in each asset can be expressed as $h_t=\pi_tw_t$,
where for $t=1,\ldots,T$
\[
w_t=\sum_{j\in J}R_{t,j}h_{t-1,j} - c_t
\]
is the {\em net wealth} of the investor at time $t$.

\subsection{Non-liability-driven investment strategies}
\label{tra_section}

In \emph{buy and hold} (B\&H) strategies the initial asset allocation $\pi_0$ is held over the subsequent time periods. To cover a nonzero claim process $c_t$, each asset is liquidated in proportion to the the initial investment. In other words, one invests
\[
h_{t,j} = 
\begin{cases}
\pi^j_0 w_0 & t=0,\\
R_{t,j}h_{t-1,j}-\pi_0^jc_t & t=1,\ldots,T,
\end{cases}
\]
units of cash in asset $j\in J$ at the beginning of the holding period starting at time $t$.

{\em Fixed proportions} (FP) is a strategy where the allocation is rebalanced at the beginning of each holding period into fixed proportions given by a constant vector $\pi\in\reals^J$ as 
\[
h_t = \pi w_t.
\]

A {\em target date fund} (TDF) is a well-known strategy in the pension industry (\cite{BodieTreussard2007}). In a TDF, the proportion invested in risky assets diminishes as a pre-determined target date approaches. We implement TDFs by adjusting the allocation between two complementary subsets $J^r$ and $J^s$ of the set of all assets $J$. Here $J^s$ consists of ``safe'' assets and $J^r$ consists of the remaining ``risky'' assets. In the simulations presented in Section 5 safe assets consist of government bonds, and risky assets comprise equities and corporate bonds.
The {\em proportional exposure} to $J^r$ at time $t$ is given by
\[
e_t = a - bt.
\]
The parameter $a$ determines the initial proportion invested in $J^r$ and $b$ defines how fast the proportion decreases in time. Choices of such $a$ and $b$ that
\[
a\ge 0\quad\text{and}\quad a-b T\ge 0.
\]
guarantees that the exposure $e_t$ in the risky assets remains nonnegative. 
A TDF can be written as
\[h_t = \pi_tw_t
\]
where the vector $\pi_t$ is adjusted to give the specified proportional exposure:
\[
\sum_{j\in J^r}\pi_{t}^j = e_t.
\]
Within $J^r$ and $J^s$ the wealth is allocated using FP rules.

\subsection{Liability-driven investment strategies}
\label{link_section}

This subsection presents strategies in which the proportions invested in different assets are connected to the development of the longevity-linked liabilities. Some of the strategies utilize the connections between mortality and financial markets observed in \cite{morfin}, while others employ, directly or indirectly, the current and forecast future cash flows of longevity-linked liabilities to determine the asset allocation.

A well-known liability-driven strategy is the {\em constant proportion portfolio insurance} (CPPI) strategy; see e.g.\ \cite{BlackJones1987, BlackPerold1992, PeroldSharpe1995}. The proportion of wealth invested in risky assets is given by
\begin{align*}
e_t &= \frac{m}{w_t}\max\{w_t-F_t,0\}\\
&= m\max\{1-\frac{F_t}{w_t},0\},
\end{align*}
where the {\em floor} $F_t$ represents the value of outstanding claims at time $t$ and the parameter $m\ge 0$ determines the fraction of the cushion (wealth over the floor) invested in risky assets. Risky and safe assets are here the same as in TDF strategies. Within $J^r$ and $J^s$ the wealth is allocated using FP rules. 
We define the floor through
\begin{align*}
F_T &= 0,\\
F_t &= (1+r)F_{t-1} - \bar c_t\quad t=0,\ldots,T,
\end{align*}
where $r$ is a deterministic discount factor and $\bar c_t$ is the median of claim amount at time~$t$. In this type of strategies, the liabilities are taken into account not only in the projected claim amounts $\bar{c}_t$ but also in the remaining wealth $w_t$, which for a given $w_0$ depends on the realized values of the claims $c_t$. 

The idea behind \emph{spread strategies} is to capture the connections between mortality and asset returns described in $\cite{morfin}$. Statistical analysis suggests that long-term increases in GDP have a positive effect on old-age mortality.
The rationale behind this is that increases in the national income are reflected in the wellbeing of the old.
On the other hand, high levels of GDP growth are connected to high term spreads. The interpretation of this connection is that interest rates reflect the future changes in the level of economic activity.

Further, when term spread is high, the yields of long-term treasury bonds are then relatively high, compared to short-term treasury bonds. Hence, relatively high yields on long-term bonds are connected with high survival probabilities of the old.
{\em Term spread strategies} aim to utilize this notion. The proportion of wealth invested in long-term treasury bonds is determined as a function of the term spread $s^T_t$ by
\[
\pi_t^L=\sigma^{a,b}(s^T_t),
\]
where
\[
\sigma^{a,b}(s) = \frac{1}{1+e^{-b(s+a)}}, 
\]
and $b>0$ and $a \in \reals$ are user-defined parameters. The remaining wealth is invested in short-term government bonds.
LISAA linkkien selitykset, ks. morfin

Correspondingly, the analysis in \cite{morfin} corroborate that high levels of GDP are connected to low credit spreads, which suggests that the low credit spreads are connected to high survival probabilities of over 50-year-olds. Simultaneously, the returns on riskier corporate bonds are relatively low compared to less risky bonds. Hence, in the case of {\em credit spread strategies} the proportional exposure to riskier corporate bonds is 
\[
\pi^{C}_t =\sigma(s^C_t) , 
\]
The remaining wealth is invested in less risky bonds.

In \emph{survival index strategies} the wealth allocated amongst assets depends on the {\em survival index} $S$ of a given population. The value of the survival index $S_t$ time $t$ is defined as the fraction of the population that survives until time $t$. The value of $S_t$ gives indication on future liabilities: the smaller the remaining number of survivors, the smaller the future cash flows are likely to be. Hence, the proportional exposure in asset $j$ at time $t$ is given by
\[
\pi^j_t = g^a(S_t),
\]
where
\[
g^a(s)= \min\{as,1\},
\]
and $a \in \reals$. The rest is invested in other assets using fixed-mix strategies. 

In \emph{wealth strategies} the proportion invested in asset $j$ depends on the proportion of initial wealth $w_t/w_0$ remaining at time $t$. The proportional exposure at time $t$ is given by
\[
\pi^j_t = g^a(w_t/w_0),
\]
where $a\in\reals$. The rest is invested in other assets. Wealth-dependent strategies resemble CPPI strategies in the sense that both define the proportions of wealth invested in various assets in terms of the present wealth. However, the wealth strategies do not depend on median liabilities like CPPI, but the liabilities are reflected only in the present level of wealth.

\section{Diversification procedure}

We now briefly recall the numerical procedure presented in \cite{fof, KoivuPennanen2010}. It is a computational method for diversifying the initial wealth $w_0$ amongst a set of simple parametric strategies called \emph{basis strategies}. 
The convex combination of feasible basis strategies is always feasible, since the optimization problem is convex. The investment strategies presented in the previous section serve as basis strategies in the numerical illustrations in Section 5. 

Consider a finite set $\{h^i\,|\,i\in I\}$ of basis strategies that invest the amount ${h^i_{t,j}}$ in asset $j$ at time $t$. The problem of finding an optimal diversification amongst the basis strategies
 can be written as 
\[
\minimize_{\alpha\in X}\quad \rho\left(\sum_{i\in I}\alpha^iw^i_T\right),
\]
where $w^i_T=\sum_{j \in J}h^i_{T,j}$ is the terminal wealth obtained by following strategy $h^i$ when starting with initial capital $w_0$, and
\[
X = \{\alpha\in\reals^I_+\,|\,\sum_{i\in I}\alpha^i=1\}.
\]
are the weights in the convex combination. In this work we employ the entropic risk measure
  \[
\rho(X)=\frac{1}{\gamma} \log E[e^{-\gamma X}],
  \]
in which case the minimization problem becomes
  \[
  \label{gp}
\minimize_{\alpha\in X}\quad \frac{1}{\gamma}\log E[e^{-\gamma \left(\sum_{i\in I}\alpha^iw^i_T\right)}].
\]
It is to be noted that our choice of the entropic risk measure is rather arbitrary and it was mainly chosen for computational convenience. Other possibilities include the {\em Conditional Value at Risk}, which is employed in an analogous setting in \cite{fof}.

Because of the convexity of $D_t$, $\sum_{i\in I}\alpha^ih^i_t \in D_t$ for $t=0,\ldots,T$. In addition, the budget constraint of the aggregate strategy $\sum_{j \in J}h_{t,j} \leq \sum_{j \in J}R_{t,j}h_{t-1,j} - c_t$ holds, if it holds for individual strategies. This is a finite-dimensional convex optimization problem, but the objective involves high-dimensional integration.

In order to solve \eqref{gp}, we form the following quadrature approximation of the objective. A finite number $N$ of return and claim scenarios $(R^k,c^k)$, $k=1,\ldots,N$ is generated over time $t=0,\ldots,T$. Here $R^k$ denotes a realization of the $|J|$-dimensional process $(R_t)_{t=1}^T$ where $R_t=(R_{t,j})_{j\in J}$.
The expectation is then approximated by 
\[
\frac{1}{N}\sum_{k=1}^N e^{-\gamma \left(\sum_{i\in I}\alpha^iw^{i,k}_T\right)} ,
\]
where $w^{i,k}_T$ is the terminal wealth in scenario $k$, obtained by following strategy $h^i$. For a more detailed description of the method, see e.g.\ \cite{fof, KoivuPennanen2010}.
 Given a realization $(R^k,c^k)$ and a strategy $h^i$, the corresponding wealth process $w^{i,k}=(w^{i,k}_ t)_{t=0}^T$ is given recursively by
\[
w_t^{i,k}=
\begin{cases}
w_0 & \text{for $t=0$},\\
\sum_{j\in J}R^k_{t,j}h^{i,k}_{t-1,j}-c^k_t & \text{for $t>0$}.
\end{cases}
\]
The resulting minimization problem is of a form that is, in principle, straightforward to solve using numerical optimization algorithms. In the numerical study below, we employ the standard SQP solver of Matlab.

\section{Numerical results}

In the following numerical illustrations, the termination date was set to $T=30$, and the cash flows $c_t$ were defined as the survival index $S_t$ of a cohort of US females aged 65 at time $t=0$. The structure of this instrument is the same as in the first longevity bond issued in 2004 by the European Investment Bank (for a more detailed description see e.g. \cite{BiffisBlake}). The asset returns $R_t$ and liabilities $c_t$ were modelled as a multivariate stochastic process as described in Appendix A. Using Latin hypercube sampling, we constructed $N=10^6$ scenarios for the numerical procedure described above. Each problem instance was generated and solved in no more than five minutes using Matlab's parallel computing on an Intel Xeon X5650 @ 2.67GHz processor.

Our aim was to investigate if liability-driven investment strategies can lead to reductions in the risk associated with a cash flow of longevity-linked liabilities.
To this end, we used two sets of basis strategies. The first set consisted of non-liability-driven basis strategies,
namely 30 FP strategies, 24 TDF strategies, and four buy and hold strategies.
The second set encompassed both the above non-liabiilty-driven and additional liability-driven basis strategies, including 15 term spread strategies, 15 credit spread strategies, 50 survival index strategies and 50 wealth strategies. We computed the optimal aggregate investment strategy and the corresponding value of the risk measure function $\rho$ for each set, using the numerical procedure of the previous section. We then proceeded to compare the optimal values of the objective $\rho$ associated with each set.
In order to discern to which extent a possible reduction in risk can be attributed to considering the liabilities,  as opposed to merely having a larger number of strategies, we also considered a portfolio optimization problem without liabilities for both sets of basis strategies. 
The optimal allocations were computed for different values of risk aversion parameters $\gamma$. The larger the parameter, the more risk averse the investor. 

Table~\ref{tab:summary} summarizes the resulting values of the objective function. We observe that as the risk aversion grows, so does the reduction in risk of the ALM problem with liabilities as the liability-driven strategies are included. This is plausible since the higher the risk aversion, the more the risk measure places importance to the fact that the asset returns conform to the liabilities. As for the optimization problem with zero liabilities, the effect of adding the liability-driven strategies was negligible and independent of the level of risk aversion. 
 
\begin{table}[!ht]
\caption{Values of objective function $\rho$.}
\label{tab:summary}
\footnotesize
\begin{center}
\begin{tabular}{c|cc|cc|cc|cc}
\hline 
&$\gamma=0.05$ & & $\gamma=0.1$ & &$\gamma=0.3$  &&$\gamma=0.5$ & \\
       & $c_t=S_t$ & $c_t=0$  & $c_t=S_t$ & $c_t=0$  & $c_t=S_t$ & $c_t=0$  & $c_t=S_t$ & $c_t=0$\\
\hline
Basis strategies &&&&&&&& \\
\hline
Non-LDI &    -27.46 &    -75.14 &     -18.64  	&    -60.82  &   -11.16  	&    -46.73  &     -9.17	&   -41.81\\
All & -27.90&  -75.14 &  -19.84	&  -60.84&  -12.40  &    -46.87	& -10.16&     -42.14\\   
&&&&&&&& \\
reduction (\%) &     1.6     &    0.006 &         6.47   &   0.04 &      11.14    &      0.3  &           10.71&      0.8\\
\hline  
\end{tabular} \\
\end{center}

\end{table}

Tables~\ref{tab:set1stra} and~\ref{tab:set2stra} show the optimal allocations to each set of the basic investment strategies and both problems for risk aversion parameter $\gamma=0.3$. Asset indexes are as indicated in Appendix A. After the liability-driven strategies were included in the optimization procedure, none of the non-liability driven strategies were included in the optimal allocation of the problem with $c_t=S_t$, whereas in the optimal allocation of the portfolio optimization problem a non-liability driven fixed proportions strategy still had the highest weight. 

\begin{table}[H]
\begin{center}
\caption{ Diversified strategy, non-liability-driven strategies, $w_0=15$, $\gamma=0.3$ }
\label{tab:set1stra}
\begin{tabular}{c|ccp{3cm}}
\hline
$c_t=S_t$&Weight (\%) & Type  & $\pi$\\
\hline
&     97.7 & FP   & \parbox{5cm}{\vspace{3pt} $\pi^2=1-0.25$ \\  $\pi^4=0.25$ \vspace{3pt}}    \\
&        2.3 & FP& \parbox{5cm}{\vspace{3pt} $\pi^2=1-0.35$ \\  $\pi^4=0.35$ \vspace{3pt}}   \\
 
\hline
$c_t=0$&Weight & Type & $\pi$  \vspace{3pt}\\
\hline
 &     59.8& FP & \parbox{5cm}{\vspace{3pt} $\pi^2=1-0.25$ \\  $\pi^4=0.25$ \vspace{3pt}}  \\
 &      40.2& FP & \parbox{5cm}{\vspace{3pt} $\pi^2=1-0.15$ \\  $\pi^4=0.15$ \vspace{3pt}}   \\
\hline
\end{tabular} \\
\end{center}
\end{table}

\begin{table}[H]
\begin{center}
\caption{ Diversified strategy, all strategies, $w_0=15$, $\gamma=0.3$ 
}
\label{tab:set2stra}
\begin{tabular}{c|ccp{2.5cm}c}
\hline
$c_t=S_t$&Weight (\%) & Type  & $\pi$ & \\
\hline
& 52.7 & Survival index & \parbox{5cm}{\vspace{3pt} $\pi^2=g^a(S_t)$ \\  $\pi^4=1-g^a(S_t)$ \vspace{3pt}}  & $a=1$   \\

 &  19.0&Wealth  & \parbox{5cm}{\vspace{3pt} $\pi^2=g^a(w_t/w_0)$ \\  $\pi^4=1-g^a(w_t/w_0)$ \vspace{3pt}}& $a=0.5$    \\

  & 13.8& Survival index & \parbox{5cm}{\vspace{3pt} $\pi^2=g^a(S_t)$ \\  $\pi^4=1-g^a(S_t)$ \vspace{3pt}}   & $a=0.75$  \\

  & 7.4&Wealth  & \parbox{5cm}{\vspace{3pt} $\pi^2=g^a(w_t/w_0)$ \\  $\pi^4=1-g^a(w_t/w_0)$ \vspace{3pt}} & $a=0.75$   \\

 & 7.1 & Term spread  & \parbox{5cm}{\vspace{3pt} $\pi^1=1-\sigma(s^T_t)^{a,b}  $ \\  $\pi^2=\sigma(s^T_t)^{a,b} $ \vspace{3pt}} & $a=-0.5, b=5$ \\
 
\hline
$c_t=0$&Weight & Type  & $\pi$ & \\
\hline
& 44.3 & FP  & \parbox{5cm}{\vspace{3pt} $\pi^2=1-0.35$ \\  $\pi^4=0.35$ \vspace{3pt}} & -  \\

& 37.6 & Term spread   & \parbox{5cm}{\vspace{3pt} $\pi^1=1-\sigma(s^T_t)^{a,b}  $ \\   $\pi^2=\sigma(s^T_t)^{a,b} $  \vspace{3pt}}& $a=-0.5, b=5$ \\

&9.6 & Survival index   & \parbox{5cm}{\vspace{3pt} $\pi^2=g^a(S_t),$ \\  $\pi^4=1-g^a(S_t),$ \vspace{3pt}} & $a=1$  \\

& 8.4&Wealth  & \parbox{5cm}{\vspace{3pt} $\pi^2=g^a(w_t/w_0)$ \\  $\pi^4=1-g^a(w_t/w_0)$ \vspace{3pt}}& $a=0.5$    \\
\hline
\end{tabular} \\
\end{center}
\end{table}

Tables~\ref{tab:nol_best} and~\ref{tab:li_best} show the best five individual strategies with the smallest risks for both problems. While all the best strategies of the problem with liabilities were liability-driven, all the best ones for the problem without liabilities were non-liability driven strategies. Note that when $c_t=0$, CPPI reduces to a fixed proportions strategy.

\begin{table}[H]
\begin{center}
\caption{ Five best basis strategies, with liabilities, $w_0=15$, $\gamma=0.3$
}
\label{tab:nol_best}
\begin{tabular}{ccp{3cm}c}
\hline
Type & Parameters & $\pi$& $\rho$\\
\hline
Survival index & $a=0.75$  & \parbox{5cm}{\vspace{3pt} $\pi^2=g^a(S_t)$ \\  $\pi^4=1-g^a(S_t)$ \vspace{3pt}}    & -11.80 \\
\hline
 Survival index & $a=1$  & \parbox{5cm}{\vspace{3pt}$\pi^2=g^a(S_t)$ \\  $\pi^4=1-g^a(S_t)$  \vspace{3pt}}   & -11.23  \\
 \hline
Wealth & $a=1$  & \parbox{5cm}{\vspace{3pt} $\pi^2=g^a(w_t/w_0)$ \\  $\pi^4=1-g^a(w_t/w_0)$ \vspace{3pt}}  & -11.22 \\
\hline
FP & -  & \parbox{5cm}{\vspace{3pt} $\pi^2=1-0.25$ \\  $\pi^4=0.25$ \vspace{3pt}}  & -11.13
  \\
\hline
CPPI &  $m=0.2, r=0.04$   & \parbox{5cm}{\vspace{3pt} $\pi^2=1-e_t$ \\  $\pi^4=e_t$ \vspace{3pt}}   & -10.89  \\
\hline
\end{tabular} \\
\end{center}
\end{table}

\begin{table}[H]
\begin{center}
\caption{ Five best basis strategies, without liabilities, $w_0=15$, $\gamma=0.3$
}
\label{tab:li_best}
\begin{tabular}{ccp{3cm}cc}
\hline
Type & Parameters & $\pi$ & $\rho$\\
\hline
 CPPI/FP &  $m=0.2, r=0.03$   & \parbox{5cm}{\vspace{3pt} $\pi^2=1-e_t$ \\  $\pi^4=e_t$ \vspace{3pt}}   & -46.62  \\
 \hline
FP & -  & \parbox{5cm}{\vspace{3pt} $\pi^2=1-0.25$ \\  $\pi^4=0.25$ \vspace{3pt}}  & -46.46 \\
  \hline
FP & -  & \parbox{5cm}{\vspace{3pt} $\pi^2=1-0.15$ \\  $\pi^4=0.15$ \vspace{3pt}}   & -46.13 \\
   \hline
TDF & $a=0.2, b=0.003$  & \parbox{5cm}{\vspace{3pt} $\pi^2=1-e_t$ \\  $\pi^4=e_t$ \vspace{3pt}}   &  -45.98 \\
 \hline
TDF &  $a=0.25, b=0.005$  & \parbox{5cm}{\vspace{3pt} $\pi^2=1-e_t$ \\  $\pi^4=e_t$ \vspace{3pt}}  &   -45.08   \\
 \hline
\end{tabular} \\
\end{center}
\end{table}

\begin{figure}[H]
\centering
\subfigure[$\gamma=0.3$, no liabilities.]
{
		\epsfig{file=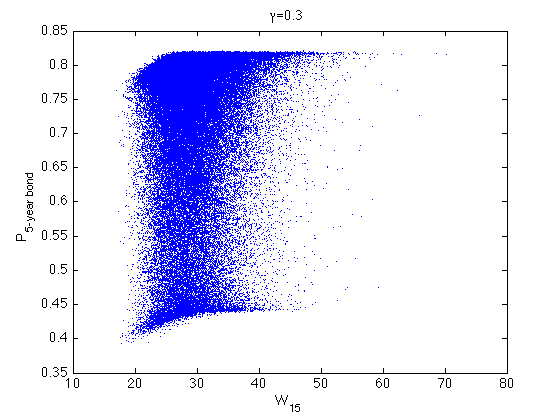,height=0.4\linewidth,width=0.47\linewidth,angle=0}
}
\subfigure[$\gamma=0.3$, with liabilities. ]
{
		\epsfig{file=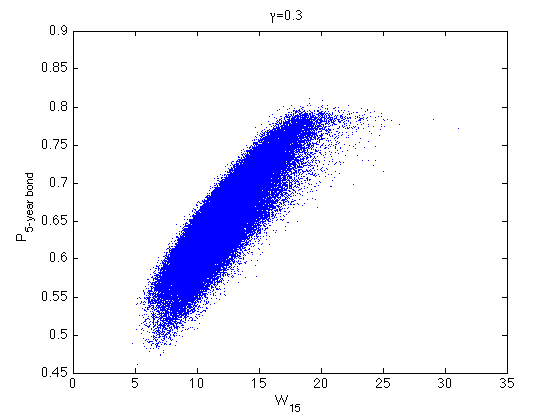,height=0.4\linewidth,width=0.47\linewidth,angle=0}
}

\caption{Proportion of wealth invested at time $t=15$ in 5-year bonds as a function of $W_{15}$. All strategies.}
\label{fig:scatters} 
\end{figure}

Figure~\ref{fig:scatters} illustrates the effect of the liability link by plotting the proportions $\pi^2$ of wealth invested in 5-year government bonds as a function of remaining wealth $w_t$ at $t=15$ in different scenarios. In the case of the ALM problem of the longevity-linked cash flow, $\pi^2$ is higher when $w_t$ is higher. For the portfolio optimization problem, however, the connection is much less clear.

\section{Conclusions}

This paper presented several liability-driven investment strategies for longevity-linked liabilities. We were able to show numerically that liability-driven investment can significantly outperform common strategies that do not take into account the liabilities. These strategies may help pension insurers and issuers of longevity-linked instruments in asset-liability management, reserving, and in underwriting new insurance contracts.

While encouraging, the results still leave substantial room for improvement. The basis strategies employed in the simulations are only an example of liability-driven strategies. Discovering and utilizing new connections between longevity-linked cash flows and asset returns would further improve the overall hedging strategy.

%


\appendix

\section{Assets and liabilities}

We consider a set $J$ of assets consisting of

\begin{enumerate}
\item US Treasury bills (1-year rate)
\item US Treasury bonds (5 -year rate) 
\item US Corporate bonds
\item US equity (S\&P total return index). 
\end{enumerate}

These are the asset classes between which the investment strategies distribute the existing wealth, and the above numbers are the indices with which the strategies are referred to. Returns on government bonds are given by the formula 
\[
R_{t}^i = \exp(Y_{t-1}^i\Delta t-D\Delta Y_{t}^i),
\]
where $Y_{t}^i$ is the yield to maturity of each bond $i \in \{{1,2}\}$ at time $t$, and $D$ is the duration. Following \cite{hylkio}, corporate bond returns are computed by 
\[
R_{t}^3 = \exp(c+(Y_{t}^3-\alpha S^C_t)\Delta t-D\Delta Y_{t}^3),
\]
where $Y_{t}^3$ is the yield to maturity of the bond, $S^C_t$ is here the credit spread between the yields of corporate bonds and longer-term government bonds $Y^2_t$, and $D$ is again the duration. Setting $c=1$ and $\alpha=1$ yields
\[
R_{t}^3 = \exp(Y_{t}^2\Delta t-D\Delta Y_{t}^3).
\]
The total return of the equity is calculated in terms of its total return index $S_t^E$,
\[
R_{t}^4 = \frac{S^E_{t}}{S^E_{t-1}}.
\]

The value of the liabilities depends on the survival index of cohort of US females aged 65 at the beginning of the observation period. The population dynamics are governed by three mortality risk factors of the mortality model presented in  \cite{AroPennanen}.

We briefly recall the stochastic model proposed in~\cite{AroPennanen}. Let $E_{x,t}$ be the number of individuals aged $[x,x+1)$ years at the beginning of year $t$ in a given population. The number of survivors $E_{x+1,t+1}$ among the $E_{x,t}$ individuals during year $[t,t+1)$ can be described by the binomial distribution:
\begin{equation}\label{eq:bin}
E_{x+1,t+1} \sim\bin(E_{x,t},p_{x,t}),
\end{equation}
where $p_{x,t}$ is the probability that an $x$ year-old individual randomly selected at the beginning of year $t$ survives until $t+1$.

The future values of $E_{t+1}$ are obtained by sampling from $\bin(E_{t},p_{x,t})$. However, as the population grows, the fraction $E_{t+1}/[E_{t}p_{x+t,t}]$ converges in distribution to constant $1$. For large populations, the population dynamics are well described by $\E=E_{x,t}p_{x,t}$, when the main uncertainty comes from unpredictable variations in the future values of $p_{x,t}$. In this work, we employ the latter approach.

 As in \cite{AroPennanen}, we model the {\em survival probabilities} $p_{x,t}$ with the formula
\begin{equation}\label{eq:logit}
p_{x,t} = \frac{\exp\left(\sum_{i=1}^n v^i_t\phi^i(x)\right)}{1+\exp(\sum_{i=1}^n v^i_t\phi^i(x))},
\end{equation}
where $\phi^i$ are user-defined {\em basis functions} and $v^i_t$ are stochastic {\em risk factors} that may vary over time. 

As in \cite{AroPennanen}, we will use the 
three piecewise linear basis functions given by
\begin{align*}
\phi^1(x) &= 
\begin{cases}
1-\frac{x-18}{32} & \text{for $x\le 50$} \\
0 & \text{for $x\ge 50$},
\end{cases}\\
\phi^2(x) &= 
\begin{cases}
\frac{1}{32}(x-18) & \text{for $x\le 50$} \\
2-\frac{x}{50} & \text{for $x\ge 50$},
\end{cases}\\
\phi^3(x) &= 
\begin{cases}
0 & \text{for $x\le 50$}\\
\frac{x}{50}-1 & \text{for $x\ge 50$}.
\end{cases}
\end{align*}
The linear combination $\sum_{i=1}^3 v^i_t\phi^i(x)$ will then be piecewise linear and continuous as a function of the age $x$. 
The risk factors $v^i_t$ now represent points on logistic survival probability curve: 
\[
v^1_t=\logit p_{18,t},\ v^2_t=\logit p_{50,t},\ v^3_t=\logit p_{100,t}.
\]

Once the basis functions $\phi^i$ are fixed, the realized values of the corresponding risk factors $v^i_t$ can be easily calculated from historical data using standard max-likelihood estimation.

As in \cite{morfin}, we model the future development of and connections between mortality risk factors and spreads with the following system of equations
\begin{align*}
\Delta v^{1}_t  &= a^{11}v^{1}_{t-1}+b^1+\varepsilon^1_t  \\
\Delta v^{2}_t &= b^2+\varepsilon^2_t  \\
\Delta v^{3}_t &= a^{33}v^{3}_{t-1}+a^{34}g_{t-1}+b^3+\varepsilon^3_t \\
\Delta g_t &= a^{45}s_{t-1}^T+a^{46}s_{t-1}^C+b^4+\varepsilon^4_t \\
\Delta s_t^T &= a^{55}s_{t-1}^T+b_5+\varepsilon^5_t \\
\Delta s_t^C &= a^{66}s_{t-1}^C+b_6+\varepsilon^6_t \\
\Delta y_t^1 &= a^{77}y^1_{t-1}+b_7+\varepsilon^7_t \\
\Delta s^E_t &=b_8+\varepsilon^8_t. \\
\end{align*}
where $v^i_t$ are mortality risk factors, $g_t$ is the natural logarithm of per capita real GDP, $s_t^T$ is the term spread between the logarithms of yields to maturity for 5-year and 1-year government bonds, and $s_t^C$ is the logarithm of the credit spread between the logarithmic yields to maturity of BAA and AAA rated corporate bonds. In addition to the risk factors included in the original model, the 1-year government bond yield $y^1_t=\textnormal{log}(Y_t^1)$ was added to enable computation of bond returns, as well as the S\&P total return index 
$s_t=\textnormal{log}(S^E_t)$ as pension plans typically invest in the stock market. 
The terms $\varepsilon_t^i$ are random variables describing the unexpected development in the risk factors.  

Once the 1-year government bond yield is known, the 5-year government bond yield can be computed by means of the term spread. Due to lack of data, we approximate the credit spread between government bonds and corporate bonds with the spread $s^C_t$ between corporate bonds of varying riskiness, obtaining the corporate bond yield.


Final year of available mortality data was 2007. Parameters of the time series model were estimated as in \cite{morfin}, with the exception that the mean reversion yields of 1-year and 5-year government bonds and corporate bonds were set to 2.5\%, 3.5\% and 4.5\%, respectively, and expected return on equity was set to 8\%.  Durations $D$ for the 1-year and 5-year Treasury bonds were 1 and 5 years, respectively, and 5 years for the corporate bonds. In the case of negative wealth, required funds were borrowed from the money market at the 1-year rate adjusted by a loan margin of 1\%.

\bibliography{mb}

\begin{thebibliography}{10}

\bibitem{AroPennanen}
H.~Aro and T.~Pennanen.
\newblock A user-friendly approach to stochastic mortality modelling.
\newblock {\em European Actuarial Journal}, 1:151--167, 2011.

\bibitem{morfin}
H.~Aro and T.~Pennanen.
\newblock Stochastic modelling of mortality and financial markets.
\newblock {\em Scandinavian Actuarial Journal}, to appear.

\bibitem{ballotta2006fair}
L.~Ballotta and S.~Haberman.
\newblock The fair valuation problem of guaranteed annuity options: The
  stochastic mortality environment case.
\newblock {\em Insurance: Mathematics and Economics}, 38(1):195--214, 2006.

\bibitem{barrieu2012understanding}
P.~Barrieu, H.~Bensusan, N.~El~Karoui, C.~Hillairet, S.~Loisel, C.~Ravanelli,
  and Y.~Salhi.
\newblock Understanding, modelling and managing longevity risk: key issues and
  main challenges.
\newblock {\em Scandinavian Actuarial journal}, 2012(3):203--231, 2012.

\bibitem{Bauer2010}
D.~Bauer, M.~Borger, and J.~Russ.
\newblock On the pricing of longevity-linked securities.
\newblock {\em Insurance: Mathematics and Economics}, 46(1):139 -- 149, 2010.

\bibitem{BiffisBlake}
E.~Biffis and D.~Blake.
\newblock Mortality-linked securities and derivatives.
\newblock {\em Pensions Institute Discussion Paper}, PI 0901, 2009.

\bibitem{BlackJones1987}
F.~Black and R.~Jones.
\newblock Simplifying portfolio insurance.
\newblock {\em Journal of Portfolio Management}, 14(1):48--51, 1987.

\bibitem{BlackPerold1992}
F.~Black and A.F. Perold.
\newblock Theory of constant proportion portfolio insurance.
\newblock {\em Journal of Economic Dynamics and Control}, 16:403--426, 1992.

\bibitem{PeroldSharpe1995}
F.~Black and A.F. Perold.
\newblock Dynamic strategies for asset allocation.
\newblock {\em Financial Analysts Journal}, 51:149--160, 1995.

\bibitem{BlakeBurrows}
D.~Blake and W.~Burrows.
\newblock Survivor bonds: Helping to hedge mortality risk.
\newblock {\em The Journal of Risk and Insurance}, 68(2):339--348, 2001.

\bibitem{BCDMortalitySecurities}
D.~Blake, A.~Cairns, and K.~Dowd.
\newblock Living with mortality: Longevity bonds and other mortality-linked
  securities.
\newblock {\em British Actuarial Journal}, 12:153--197(45), 2006.

\bibitem{Blake2012}
D.~Blake, A.~Cairns, K.~Dowd, and G.~Coughlan.
\newblock Longevity hedge effectiveness: A decomposition.
\newblock {\em Quantitative Finance}, 2012, forthcoming.

\bibitem{BCDM2006}
D.~Blake, A.~Cairns, K.~Dowd, and R.~MacMinn.
\newblock Longevity bonds: Financial engineering, valuation, and hedging.
\newblock {\em The Journal of Risk and Insurance}, 73(4):pp. 647--672, 2006.

\bibitem{BodieTreussard2007}
Z.~Bodie and J.~Treussard.
\newblock Making investment choices as simple as possible, but not simpler.
\newblock {\em Financial Analysts Journal}, 63(3):42--47, 2007.

\bibitem{Cairns2011a}
A.~Cairns.
\newblock Modelling and management of longevity risk: approximations to
  survival functions and dynamic hedging.
\newblock {\em Insurance: Mathematics and Economics}, 49, 2011.

\bibitem{PricingDeath}
A.~Cairns, D.~Blake, and K.~Dowd.
\newblock Pricing death: Frameworks for the valuation and securitization of
  mortality risk.
\newblock {\em ASTIN bulletin}, 36(1):pp. 79--120, 2006.

\bibitem{cox2007natural}
S.~Cox and Y.~Lin.
\newblock Natural hedging of life and annuity mortality risks.
\newblock {\em North American Actuarial Journal}, 11(3):1--15, 2007.

\bibitem{DMM2008}
M.~Dahl, M.~Melchior, and T.~Møller.
\newblock On systematic mortality risk and risk-minimization with survivor
  swaps.
\newblock {\em Scandinavian Actuarial Journal}, (2/3):114 -- 146, 2008.

\bibitem{ds6}
F.~Delbaen and W.~Schachermayer.
\newblock {\em The Mathematics of Arbitrage}.
\newblock Springer Finance. Springer-Verlag, Berlin Heidelberg, 2006.

\bibitem{DowdSurvivorSwaps}
K.~Dowd, D.~Blake, A.~Cairns, and P.~Dawson.
\newblock Survivor swaps.
\newblock {\em The Journal of Risk and Insurance}, 73(1):1--17, 2006.

\bibitem{mdfrm}
K.~Dowd, A.~Cairns, and D.~Blake.
\newblock Mortality-dependent financial risk measures.
\newblock {\em Insurance: Mathematics and Economics}, 38(3):427 -- 440, 2006.

\bibitem{FS}
H.~F\"ollmer and A.~Schied.
\newblock {\em Stochastic Finance}.
\newblock Walter de Gruyter, Berlin, 2004.

\bibitem{fof}
P.~Hilli, M.~Koivu, and T.~Pennanen.
\newblock Optimal construction of a fund of funds.
\newblock {\em European Actuarial Journal}, 1:345-359, 2011.

\bibitem{hylkio}
M.~Koivu and T.~Pennanen.
\newblock Two-factor models for index linked bond portfolios.
\newblock {\em Manuscript}.

\bibitem{KoivuPennanen2010}
M.~Koivu and T.~Pennanen.
\newblock Galerkin methods in dynamic stochastic programming.
\newblock {\em Optimization}, 59:339--354, 2010.

\bibitem{LiHardy2011}
J.S.-H. Li and M.R. Hardy.
\newblock Measuring basis risk in longevity hedges.
\newblock {\em North American Actuarial Journal}, 15, 2011.

\bibitem{LinCoxMBonds}
Y.~Lin and S.~H. Cox.
\newblock Securitization of mortality risks in life annuities.
\newblock {\em The Journal of Risk and Insurance}, 72(2):227--252, 2005.

\bibitem{pen13}
T.~Pennanen.
\newblock Risk management and contingent claim valuation in illiquid markets.
\newblock {\em Manuscript}.

\bibitem{NaturalWang2010}
J.~L. Wang, H.C. Huang, S.~S. Yang, and J.~T. Tsai.
\newblock An optimal product mix for hedging longevity risk in life insurance
  companies: The immunization theory approach.
\newblock {\em Journal of Risk and Insurance}, 77(2):473 -- 497, 2010.

\bibitem{wilkie2003reserving}
A.~Wilkie, H.~Waters, and S.~Yang.
\newblock Reserving, pricing and hedging for policies with guaranteed annuity
  options.
\newblock {\em British Actuarial Journal}, 9(02):263--391, 2003.

\end{thebibliography}
\bibliographystyle{plain}

\end{document}